\documentclass[twocolumn,final]{article}
\usepackage{preprint}
\usepackage{wrapfig}
\usepackage{graphicx}
\usepackage[hidelinks]{hyperref}
\usepackage{pbalance}

\usepackage{ulem}
\usepackage{tabularray}
\usepackage{multirow}
\usepackage[bottom]{footmisc}
\usepackage{cite}
\usepackage{amsmath,amssymb,amsfonts}
\usepackage{cleveref}
\usepackage{algorithmic}
\usepackage{textcomp}

\usepackage{xcolor}

\usepackage[frozencache,cachedir=minted-cache]{minted}
\def\BibTeX{{\rm B\kern-.05em{\sc i\kern-.025em b}\kern-.08em
    T\kern-.1667em\lower.7ex\hbox{E}\kern-.125emX}}
\usepackage{tikz}
\newcommand{\circled}[1]{
    \tikz[baseline=(char.base)]{
            \node[shape=circle,draw,inner sep=0.75pt] (char) {#1};
    }
}

\newcommand{\name}{\textsc{Oxn}}

\title{\name{} - Automated Observability Assessments for Cloud-Native Applications\thanks{}}

\usepackage{authblk}

\author[1]{Maria C. Borges}
\author[1]{Joshua Bauer}
\author[1]{Sebastian Werner} 
\affil[1]{Information Systems Engineering,
Technische Universität Berlin, Germany}

\begin{document}
\twocolumn[\begin{@twocolumnfalse}
\maketitle
\pagestyle{plain}

\begin{abstract}
Observability is important to ensure the reliability of microservice applications. These applications are often prone to failures, since they have many independent services deployed on heterogeneous environments. When employed ``correctly", observability can help developers identify and troubleshoot faults quickly. However, instrumenting and configuring the observability of a microservice application is not trivial but tool-dependent and tied to costs. 
Practitioners need to understand observability-related trade-offs in order to weigh between different observability design alternatives. 
Still, these architectural design decisions are not supported by systematic methods and typically just rely on ``professional intuition".  

To assess observability design trade-offs with concrete evidence, we advocate for conducting experiments that compare various design alternatives. Achieving a systematic and repeatable experiment process necessitates automation. We present a proof-of-concept implementation of an experiment tool --- \textbf{O}bservability e\textbf{X}periment e\textbf{N}gine (OXN). \name{} is able to inject arbitrary faults into an application, similar to Chaos Engineering, but also possesses the unique capability to modify the observability configuration, allowing for the straightforward assessment of design decisions that were previously left unexplored. 

\end{abstract}
\vspace{0.5cm}

\end{@twocolumnfalse}]
{
  \renewcommand{\thefootnote}
    {\fnsymbol{footnote}}
  \footnotetext[1]{Poster available under: \url{doi.org/10.5281/zenodo.12734079}
}
\renewcommand{\thefootnote}{\arabic{footnote}}
\section{Introduction}Modern cloud-native applications benefit from the agility and scalability of the microservice architectural style, but their distributed nature also presents significant challenges with regards to reliability.
In such dynamic and evolving environments, not everything can be tested upfront, so these applications are typically more prone to faults, which can be unpredictable and hard to diagnose \cite{Zhang_MicroserviceSurvey_ThresholdingHard_2019,Niedermaier_ObservabilityInterviewStudy_2019}. 
To address this challenge, developers can instrument their applications with monitoring, tracing and logging. Observability, when employed correctly, helps improve reliability  
by providing developers with the necessary information to quickly detect and resolve faults.

Still, designing the observability of a microservice application is not trivial. 
It raises a number of important considerations and decisions that go beyond what tool to select.
Particularly, developers must consider the instrumentation, as some aspects can be covered by the built-in instrumentation of different components, while others necessitate custom instrumentation logic.
Furthermore, configuring the parameters of each tool correctly is also important; otherwise, the observability data can mask issues and prolong fault diagnosis \cite{Borges_TracingServerless_2021}. 
Additionally, it is important to remember that setting up and maintaining observability infrastructure also comes with its own costs and overhead, which need to be weighed against the benefits provided by the observability.
All these design decisions don't happen just once, but instead need to the considered continuously, with every system change.

Given the large design space for observability, one would expect there to be established methods to inform these design decisions. 
However, in the past, observability design decisions have typically not been supported by evidence or systematic methods. 
Instead, many decisions are either impulsive, in response to an acute problem \cite{Niedermaier_ObservabilityInterviewStudy_2019}, or rely on individual experience and ``professional intuition" of developers \cite{Zhang_MicroserviceSurvey_ThresholdingHard_2019,Vale_MicroserviceQualityMeasurementIndustrySurvey_2022}.
Unfortunately, it is very hard for developers to obtain factual information on whether their observability setup is fit for purpose. 

Experimentation and benchmarking serve as established methods for assessing the non-functional properties of various systems. 
Traditionally, they have been used to evaluate conventional qualities like performance and scalability. 
More recently, developers have extended these methods to evaluate the reliability and resilience of applications, a practice known as Chaos Engineering \cite{Basiri_ChaosEngineeringPrinciples_2016}. 
While some sporadic experiments have been conducted on the observability of certain systems in the past\cite{Ahmed_EffectivenessOfAPM_2016}, experimentation has yet to become a common practice in the observability decision making process. 
Nevertheless, we believe that with the right tooling we can reduce the effort involved in observability assessment, so that experimentation can become a viable approach in support of observability decision making.  

This paper introduces \name{} ({the \textbf{O}bservability E\textbf{x}periment E\textbf{n}gine}) -- a tool designed to facilitate systematic, reproducible, and comparative assessment of observability design decisions.
 In the following, we show how our tool builds upon related research and industry efforts, present its architecture, demonstrate its use and discuss future directions.
\label{sec:intro}
\section{Related Research Efforts}Observability has been evaluated by means of experimentation several times in the past, though these assessments have predominantly focused on the costs associated with observability. For instance, Reichelt et al. \cite{Reichelt_OverheadOpenTelemetryEtc} conducted a concrete benchmark of observability instrumentation. Their study introduced a tool for continuous measurement of overhead of popular instrumentation libraries.
More recently, Dinga et al. \cite{Dinga_EnergyEfficiencyOfMonitoring_2023} experimentally investigated the energy efficiency of different observability tools. 
In another experiment-based study, Ahmed et al. \cite{Ahmed_EffectivenessOfAPM_2016} explored the effectiveness of four monitoring tools in identifying performance regressions but assumed the default configuration for every tool, therefore ignoring instrumentation and configuration decisions. While all of the three papers investigate an important slice of the observability design landscape, a comprehensive exploration of design decisions is still missing.

Besides observability, our work also draws heavily upon the discipline of chaos engineering. Chaos Engineering provides a method  \cite{Basiri_ChaosEngineeringPrinciples_2016} and tools \cite{Heorhiadi_GremlinChaosEngineering_2016,Meiklejohn_FillibusterChaosEngineering_2021,Zhang_PhoebeChaosEngineering_2022} for carrying out resilience experiments. Here, faults are injected into a running microservice system to test the systems ability to withstand faults. However, good observability of the system is implicitly assumed for the approach to work, since it typically  relies on data from tailored, product-centric metrics to evaluate system behavior against expectations \cite{Basiri_ChaosEngineeringPrinciples_2016}.

Given the lack of dedicated tooling for observability experimentation, we developed \name{}. \label{sec:relwork}
\section{OXN: Observability Experiment Engine}\subsection{Overview and Key Design Features}
We propose an empirical and systematic approach to navigate the large observability design space: observability experiments. Experimentation is an established approach for assessing other non-functional properties of cloud-native applications \cite{Klems_phd_2016} and has also been used to assess observability in the past \cite{Ahmed_EffectivenessOfAPM_2016}. However, such experimentation comes with distinct challenges:

\textbf{C1:} Experiments can be difficult to reproduce, re-run and validate, \textbf{C2:} Experiments sometimes struggle to mimic realistic application scenarios, compromising the relevance of the results \textbf{C3:} Experiments involve many moving parts and can be difficult to setup and manage, especially for complex cloud-native stacks, \textbf{C4:} Measuring observability effectiveness is not as straightforward as measuring performance and therefore requires new measurement and reporting approaches. 

Nevertheless, we believe that with the right tooling we can overcome some challenges and reduce the barriers associated with observability experiments. 
We translated these challenges into the following design features of \name{}:
\subsubsection*{\textbf{D1 - Systematic experiment specification}} \name{} follows the design principles of cloud benchmarking \cite{2017-Bermbach-Book-CloudServiceBenchmarking, Silva_cloudbench_2013, Klems_phd_2016} and thus particularly ensures portable and repeatable experiments. Experiments are defined as \textsc{yaml}-based configuration files, which allows them to be shared, versioned and repeated. The experiment configuration is self-contained, meaning it includes all the necessary information to execute or reproduce an experiment. 

\subsubsection*{\textbf{D2 - Custom load curves and extensible treatment library}}
A varying load can directly influence the data points received by the observability backend. Workload is therefore another aspect that needs to be considered in the experiment specification. Here, we allow users to specify complex load patterns by providing key-value pairs that define load at different stages of an experiment. This fine-grained control over the load generation essentially allows users to construct any arbitrary load shape. When modeled after an application's production workload, this feature can therefore provide loads that mirror real application scenarios.

Treatments are controlled changes to the system under experiment. We distinguish between fault treatments and instrumentation treatments. \name{} already provides a core set of treatments out-of-the-box, which cover some basic use-cases and serve as proof-of-concept. Behind these treatments lies a common interface that can be leveraged by practitioners to implement custom treatments. This extensible approach encourages the recreation of real fault scenarios, for example as part of postmortem culture \cite{Beyer_GoogleSRE_2016}.

\subsubsection*{\textbf{D3 - Automated experiment setup and execution}} To foster widespread use of observability experiments, \name{} automates every step of the experiment process in a straightforward manner, from system under experiment (SUE) setup to data collection, processing and reporting. 

\subsubsection*{\textbf{D4 - Fault observability measurement approach}} To arrive at explicit metrics for observability, we restrict our scope to fault observability \cite{Borges_TracingServerless_2021, Borges_InformedObservabilityDecisions_2024}. Essentially, we feed the observability data generated during the experiment through a fault detection mechanism and report if the faults  were detected in any of the observed metrics. A simple fault detection mechanism is already included in \name{}, but practitioners can also specify a custom mechanism against an interface. 

\begin{figure}
    \centering
    \includegraphics[width=1\linewidth]{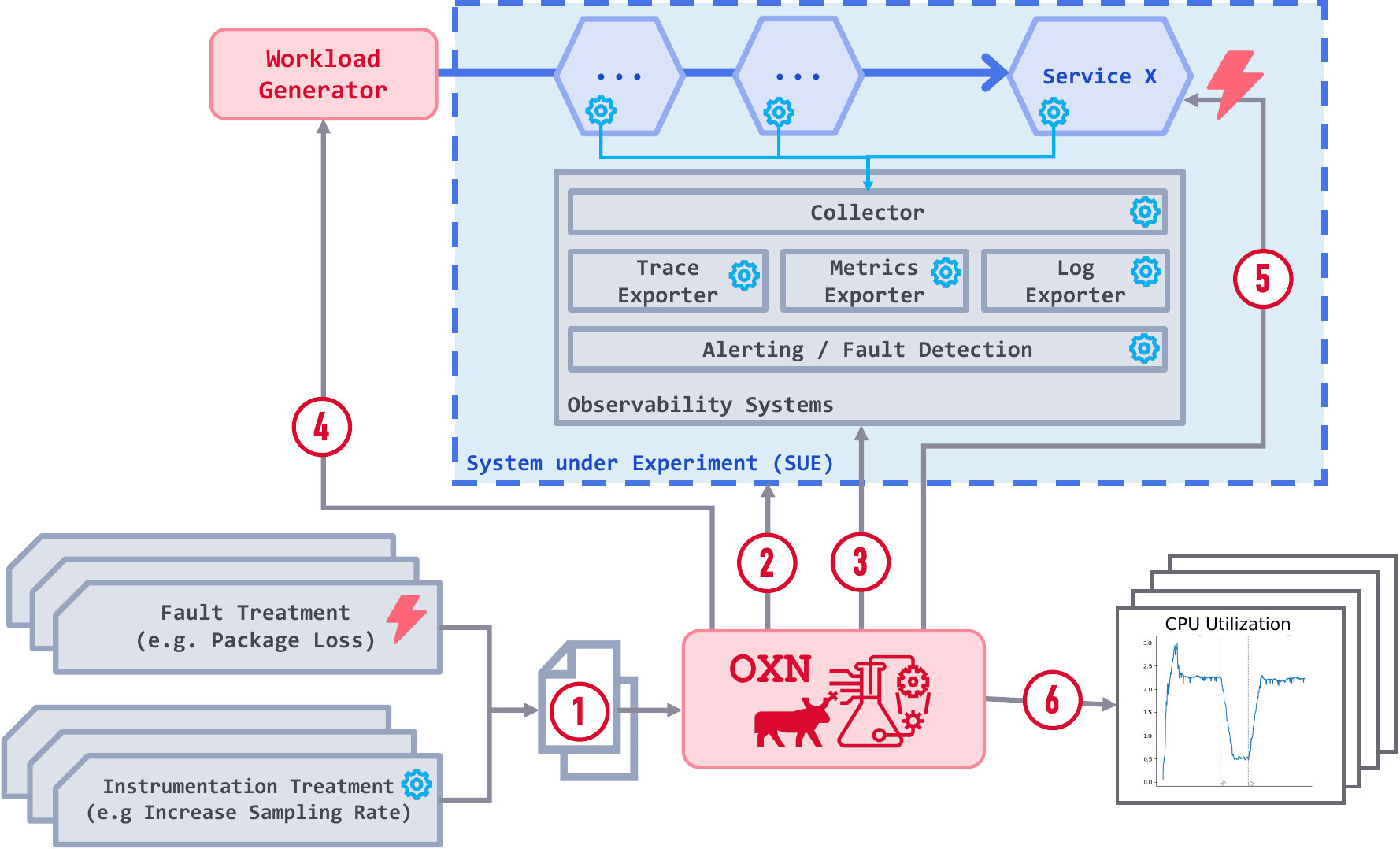}
    \caption{\name{}'s operation for performing automated observability assessments.} 
    \label{fig:usage}
\end{figure}
\subsection{Mode of Operation}
\Cref{fig:usage} illustrates the process of automated observability assessments with \name{}. To begin, the practitioner selects or creates a set of experiments \circled{1}(see Listing \ref{listing:experimentyaml} for more details).
Then \name{} deploys the SUE \circled{2}, which needs to have its deployment specified via IaC, e.g., through Docker-Compose, Kubernetes, or Terraform. Afterwards, \name{} applies all the instrumentation treatments included in the experiment set \circled{3}. Once the SUE is up and running, \name{} starts the workload generator \circled{4}, and soon after can start sequentially applying the fault treatments. Throughout the entire process \name{} collects data from the observability system \circled{5} and generates a report \circled{6}for the practitioner to review. This report is used by a jupyter notebook to assess whether the faults were detectable and is complemented by plots of the metrics.  

Practitioners can use \name{} as a standalone tool to sporadically assess the suitability of their current observability configuration against given fault scenarios. Alternatively, \name{} can also be integrated into a release pipeline, to continuously assess tradeoffs between fault visibility and observability costs, for example, to satisfy SLOs for system reliability. \label{sec:oxn}
\section{Architecture and Implementation}\begin{figure*}[h]
    \centering
    \includegraphics[width=0.9\textwidth]{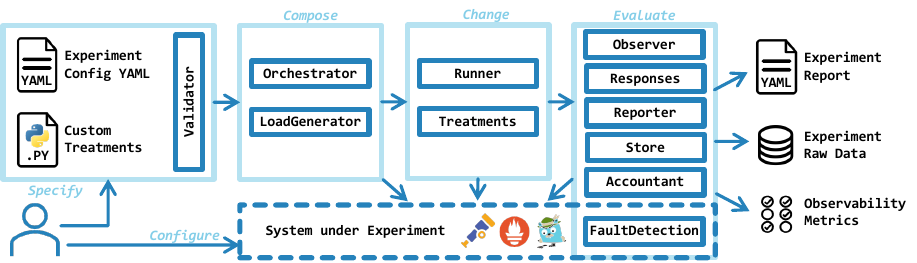}
    \caption{System architecture and implementation of \name{}.}
    \label{fig:oxn-architecture}
\end{figure*}
The design of \name{}\footnote[1]{Available at \url{https://github.com/nymphbox/oxn}
} is modular, decoupled, and extensible. \Cref{fig:oxn-architecture} shows a simplified view of the architecture, which also reflects the steps of the experimentation process. This section describes notable artifacts and components of \name{}.

\begin{table}[t]
    \caption{Different treatments implemented in \textsc{oxn}} \label{table:instrumentation-treatment-library}\label{table:fault-treatment-library}
    \resizebox{\linewidth}{!}{
        \newcounter{treatmentsFoodnoteCounter}
\setcounter{treatmentsFoodnoteCounter}{\value{footnote}} \centering
\newcommand{\rightIndent}{\hspace{0.2cm}}
\begin{tabular}{lp{0.7\linewidth}l}
    Name & Purpose & Tool \\ \hline
    \multicolumn{3}{l}{\textbf{Fault Treatment:}} \\
    \rightIndent{}Pause & Simulates an unresponsive service by suspending all processes in a service & \textsc{Docker} \\
    \rightIndent{}Kill & Simulates a service crash & \textsc{Docker} \\
    \rightIndent{}NetworkDelay & Injects network delay on an interface & \textsc{tc\footnotemark} \\
    \rightIndent{}PacketLoss & Injects packet loss on an interface & \textsc{tc} \\
    \rightIndent{}PacketCorruption & Injects packet corruption on an interface & \textsc{tc} \\
    \rightIndent{}Stress & Simulates resource exhaustion by injecting stressors & \textsc{stress-ng\footnotemark} \\ \hline
    \multicolumn{3}{l}{\textbf{Instrumentation Treatment:}} \\
    \rightIndent{}MetricSamplingRate & Changes the sampling rate for metrics & \textsc{Collector} \\
    \rightIndent{}TracingSamplingStrategy & Samples traces based on a given Strategy & \textsc{Collector} \\
    \rightIndent{}TracingSamplingRate & Samples traces based on a given sampling rate & \textsc{Collector}
\end{tabular}

    }
\vspace{-1em}
\end{table}
\stepcounter{treatmentsFoodnoteCounter}
\footnotetext[\value{treatmentsFoodnoteCounter}]{\href{https://man7.org/linux/man-pages/man8/tc.8.html}{linux traffic control}}
\stepcounter{treatmentsFoodnoteCounter}
\footnotetext[\value{treatmentsFoodnoteCounter}]{\href{https://manpages.ubuntu.com/manpages/bionic/man1/stress-ng.1.html}{linux kernel load and stress testing tool}}

\subsection*{Experiment Config YAML}
An observability experiment can be defined through a syntactically valid YAML file, which provides a declarative and machine-readable specification for the experiment. 
\Cref{listing:experimentyaml} shows an example.
The experiment specification consists of four sections. 

\begin{listing}[b!]
\begin{minted}[breaklines, fontsize=\fontsize{6}{8}]{yaml}
experiment:
  responses:
    - recommendations_per_min:
        type: metric
        metric_name: increase(app_recommendations_counter_total[1m])
        left_window: 240s 
        right_window: 240s
        step: 1
  treatments:
    # change of SUE configuration
    - change_metric_interval:
        action: otel_metrics_interval
        params: {
          service_name: recommendation-service,
          export_interval_ms: 1000
        }
    # fault to be injected
    - package_loss_treatment:
        action: loss
        params: {
          service_name: recommendation-service,
          duration: 120s,
          loss_percentage: 50%,
          interface: eth0,
        }
  sue:
    compose: opentelemetry-demo/docker-compose.yml
    exclude: [loadgenerator]
  loadgen:
    run_time: 10m
    stages: 
    - {duration: 600, users: 50, spawn_rate: 25}
    tasks:
    # specific requests to perform per spawned user
    - { endpoint: /, verb: get, params: { } }
\end{minted}
\caption{Example of an \name{} \textit{Experiment Config}-file.}
\label{listing:experimentyaml}
\end{listing}    

The \textbf{\texttt{responses}} section defines what particular metrics and traces generated by the system are of interest for the assessment. Users can choose relevant metrics and declare metric-specific parameters through key-value pairs. It is also here where the user can define the time window for the observations.

Next, users define a list of so-called \textbf{\texttt{treatments}}. We distinguish between fault treatments and instrumentation treatments. Instrumentation treatments allow users to easily change observability configuration without having to tinker with the SUE code and are executed during compilation. Fault treatments, in turn, are applied at runtime and change the SUE by means of dynamic fault injection. Users can choose from \name{}'s library of included treatments (see also Table \ref{table:fault-treatment-library}), then specify treatment-specific parameters like fault duration.

After, in the \textbf{\texttt{sue}} section, users can define directives that relate to the orchestration of the system under experiment. Here, \name{} for now relies on a docker-compose file for the SUE and users also have the option to exclude or include certain services if they only wish to experiment on a subset of the original SUE. In Listing \ref{listing:experimentyaml}, for example, we exclude the default load generator packaged with the application to instead define a custom workload next.

Lastly, in the \textbf{\texttt{loadgen}} section, users can describe their workload by defining basic tasks to be executed by the load generator. Through staging, they can also increase or decrease the load throughout the experiment duration.

\subsection*{Orchestrator and Load Generator}
We provide an implementation of an orchestrator that relies on docker to compose arbitrary systems under experiment. 
For load generation, we make use of 
Locust\footnote{\url{https://locust.io}}.

\subsection*{Observer, Responses, Store, Reporter and FaultDetection}
The observer module provides methods that allow the runner to pass experiment information, e.g. execution start and end timestamps, to the response variables for labelling purposes. 

The responses module implements metric and trace response variables from the base response variable model. These implementations abstract over data collection and data labeling details and utilize the Prometheus and Jaeger interfaces to read response data. They also implement different labeling strategies depending on the data type.

The store module is a simple abstraction over HDF, a binary data format that is readable by most standard data analysis solutions. We chose HDF for its simplicity and fast read and write speeds. It also allows the annotation of arbitrary metadata, which facilitates the user when working with data generated by \name{}.

 A reporter in \name{} can generate a machine-readable experiment report with a quick overview of treatments and the  location of the respective observability data. This data is then analyzed in a jupyter notebook provided with the tool. It generates plots and shows the effectiveness of the implemented observability by calculating observability metrics \cite{Borges_InformedObservabilityDecisions_2024}. For these metrics, we offer an extensible interface, enabling developers to seamlessly add their own fault detection mechanisms, which could range from threshold-based alerting techniques to more advanced anomaly detection models.

 \label{sec:architecture}
\section{Conclusion}\label{sec:concl}
With \name{}, we present the first tool of its kind that allows for on-demand assessment of observability through experimentation. It shows practitioners the effectiveness of their observability and reveals observability-related trade-offs. With comparative assessments, practitioners can justify design decisions when it comes to configuring observability tools, instrumenting services, or defining thresholds. Moreover, \name{} can be used to generate reproducible training data for anomaly detection or for any system that aims to recognize faults in complex cloud-native microservice deployments.

For future work, we aim to improve the applications of \name{} by adapting it to other deployment frameworks such as Kubernetes and by improving fault injection and treatments. Besides that, we hope that other practitioners start implementing and sharing fault scenarios and treatment options to enable the community as a whole to create more resilient cloud native systems. 

\section*{Acknowledgements}

\small \noindent Funded by the European Union (TEADAL, 101070186). Views and opinions expressed are however those of the author(s) only and do not necessarily reflect those of the European Union. Neither the European Union nor the granting authority can be held responsible for them.

\bibliographystyle{ieeetr}
\bibliography{refs}

\end{document}